\documentclass[a4paper,11pt]{article}

\usepackage{amssymb,amsmath,amsfonts,epsfig}
\usepackage{tikz}
\usetikzlibrary{shapes, arrows, shadows}

\def\be{\begin{equation}}
\def\ee{\end{equation}}
\def\bea{\begin{eqnarray}}
\def\eea{\end{eqnarray}}

\newcommand{\im}{\mathrm{Im\,}}

\newcommand{\comment}[1]{}

\let\oldsqrt\sqrt
\def\sqrt{\mathpalette\DHLhksqrt}
\def\DHLhksqrt#1#2{%
\setbox0=\hbox{$#1\oldsqrt{#2\,}$}\dimen0=\ht0
\advance\dimen0-0.2\ht0
\setbox2=\hbox{\vrule height\ht0 depth -\dimen0}%
{\box0\lower0.4pt\box2}}

\begin{document}
\begin{titlepage}
\date{today}       \hfill
{\raggedleft IPMU14-0169}
\begin{center}
\vskip .5in
{\Large \bf  Entropy of conformal perturbation defects}\\
 
\vskip .250in

\vskip .5in
{\large Anatoly Konechny${^{1,2,\sharp}}$ and Cornelius Schmidt-Colinet${}^{3,\dagger}$ }

\vskip 0.5cm
{\it ${}^{1}$School of Mathematical and Computer Sciences, The University of Heriot-Watt\\
Edinburgh EH14 4AS, United Kingdom\\[10pt]

${}^{2}$Maxwell Institute for Mathematical Sciences\\
Edinburgh, United Kingdom\\[10pt]

${}^{3}$Kavli Institute for the Physics and Mathematics of the Universe (WPI)\\ 
Todai Institutes for Advanced Study, The University of Tokyo\\
Kashiwa, Chiba 277-8583, Japan}
\end{center}

\vskip .5in

\begin{abstract} \large
We consider perturbation defects obtained by perturbing a 2D conformal  field theory (CFT) by a 
relevant operator on a half-plane. If the perturbed bulk theory flows to an infrared fixed point 
described by another CFT, the defect flows to a conformal defect between the ultraviolet and 
infrared fixed point CFTs. For short bulk renormalization group flows connecting two fixed points 
which are close in theory space we find a universal perturbative formula for the boundary entropy of the 
corresponding conformal perturbation defect. We compare the value of the boundary entropy that our
formula gives for the flows between nearby Virasoro minimal models ${\cal M}_m$ with the boundary entropy
of the defect constructed by Gaiotto in \cite{Gaiotto} and find a match at the first two orders in 
the $1/m$ expansion. 
\end{abstract}
\vfill
{\small ${}^\sharp$  {\tt a.konechny@hw.ac.uk}\\
${}^\dagger$ {\tt cornelius.schmidt-colinet@ipmu.jp}}

\end{titlepage}
\large 

\section{Introduction}
We are interested in renormalization group flows of two-dimensional quantum field theories. The 
flow originates from an ultraviolet fixed point and flows into an infrared one (which may be trivial). 
The  fixed points enjoy the infinite-dimensional conformal symmetry which is a powerful tool in solving them \cite{BPZ}. 
Besides the traditional objects of interest --- local operators and their correlation functions ---
boundary conditions and conformal defects (interfaces between 2d CFTs) have received much 
attention. Such objects are not only interesting in their own right, but have proved to be an 
important tool in classifying and solving CFTs (see {\it e.g.} \cite{Schweigert_etal} and 
references therein).  In \cite{FQ,BR} an interesting idea was put forward which associates
a defect between UV and IR fixed points, described by ${\rm CFT}_{\rm UV} $ and 
${\rm CFT}_{\rm IR}$ respectively,  with a renormalization group flow connecting them. If the flow 
is triggered by  a perturbation $\Delta S = \int\!\! d^2 x \, \lambda^i \phi_i(x) $ of the UV fixed point, 
one can consider this perturbation and the subsequent RG flow on a half plane 
$x_1>0$.

\begin{center}
\begin{tikzpicture}[scale=1.4]
\filldraw[fill=blue!10!white, draw=white] (0, 0.018) rectangle (3,3); 
\filldraw[fill=red!10!white, draw=white] (-3,0) rectangle (-0.018,3);
\draw[very thick] (0,0) -- (0,3);
\draw (-2,1.5) node {$\displaystyle{e^{\int\!\! \lambda^i\phi_i(x)d^2 x}}$};
\draw (-2,0.5) node {{\small Perturbed }};
\draw (1.5, 0.5) node {{\small Unperturbed }} ;
\end{tikzpicture}
\end{center}

It may happen that new divergences arise when the insertions of $\phi_i(x)$ collide at the 
boundary $x_1=0$ of the perturbed region. Renormalization will then require the introduction of
new boundary couplings, which will flow together with the bulk couplings $\lambda^i$. We can fold 
the bulk theory along the $x_1=0$ line and look at the resulting flow on a half plane $x_1\ge 0$ as 
a coupled bulk plus boundary flow (such a folding trick \cite{Oshikawa-Affleck} is customary in defect theory). 
On the interior of a half plane the bulk flow connects the tensor product of two copies of the UV theory 
${\rm CFT}_{\rm UV}\otimes {\rm CFT}_{\rm UV}$ to ${\rm CFT}_{\rm UV}\otimes {\rm CFT}_{\rm IR}$, 
while the induced boundary flow connects the trivial conformal boundary condition in 
${\rm CFT}_{\rm UV}\otimes {\rm CFT}_{\rm UV}$ to some conformal boundary condition in 
${\rm CFT}_{\rm UV}\otimes {\rm CFT}_{\rm IR}$. Unfolding the last object we obtain a conformal
defect between ${\rm CFT}_{\rm UV}$ and ${\rm CFT}_{IR}$ associated with the bulk flow.
We propose to  call such defects ``conformal perturbation defects". In this paper for brevity 
we will often omit the word ``conformal".   The idea 
that one has a conformal object, which can be handled using the powerful algebraic techniques 
of CFTs, that carries information about an RG flow is very appealing. 
The important question seems to be --- exactly what information do perturbation  defects carry about 
the bulk flows?

Given a conformal boundary condition $|B\rangle$ in ${\rm CFT}_{\rm UV}$, a bulk flow will 
typically induce some boundary RG flow that will bring $|B\rangle_{\rm UV}$ to some conformal 
boundary condition $|B'\rangle_{\rm IR}$ in ${\rm CFT}_{\rm UV}$ \cite{Gaberdiel_etal}. 
It was suggested in \cite{BR} that $|B'\rangle_{\rm IR}$ can be obtained by fusing the 
perturbation defect associated with the bulk flow with  $|B\rangle$. Concrete RG flows between 
supersymmetric $N=2$ minimal models were studied in \cite{BR} and the proposed fusion rule was 
shown to hold. 

In \cite{Gaiotto} the general idea of associating a conformal defect to bulk RG flows was 
approached from a different angle. Any conformal defect between ${\rm CFT}_{\rm UV}$ and 
${\rm CFT}_{\rm IR}$ gives a pairing of local operators in the two theories. The pairing between 
operators $\phi^{\rm UV}$ and $\phi^{\rm IR}$  is obtained by inserting  $\phi^{\rm UV}$ at the 
origin, surrounding it with the defect placed on a circle at some radius, and inserting 
$\phi^{\rm IR}$ at infinity. In RG flows the operators in ${\rm CFT}_{\rm IR}$ can be represented 
in terms of operators in  ${\rm CFT}_{\rm UV}$. Picking a basis in each theory we have
\be \label{bij}
\phi^{\rm IR}_i= b_{i}^{\;j} \phi^{\rm UV}_{j} \, . 
\ee
The coefficients $b_i^{\;j}$ essentially can be computed from the RG mixing matrices driven to the IR 
fixed point. It was conjectured in \cite{Gaiotto} that special conformal defects (RG domain walls) 
exist for which the natural pairing between $\phi^{\rm IR}_i$ and $\phi^{\rm UV}_j$ is equal to the 
RG coefficients $b_i^{\;j}$. For the RG flows between two neighbouring  minimal models described in 
\cite{Zamolodchikov} a candidate defect which is conjectured to have this property was constructed in 
\cite{Gaiotto}. We discuss this defect in some more detail in Sections \ref{G_defect} and \ref{conclusion}. 
Here we would like to note that it is not clear whether such a defect is unique in view of the scheme 
dependence of the coefficients $b_i^{\;j}$. It is also not clear what the relation between the RG domain 
walls of \cite{Gaiotto} and the perturbation defects of \cite{BR} is in general. 

In this paper we consider perturbation defects, associated to ``short" RG  flows in the bulk. Such 
a flow  is triggered by perturbing the UV CFT by a nearly marginal operator $\phi$ with 
scaling dimension $\Delta=2-\delta$, where $\delta \ll 1$ can be used as a dimensionless
perturbation parameter. We assume that the OPE of $\phi$ with itself,
\be\label{OPE}
\phi(x) \phi(0) = \frac{1}{|x|^{2\Delta}} +  \frac{C}{|x|^{\Delta}} \phi(0)  
+ \mbox{irrelevant fields}\,,
\ee 
does not contain any other relevant operators besides the identity and $\phi$ itself. If $\lambda$ 
is the renormalised coupling constant corresponding to the $\phi$-perturbation, the beta function 
at quadratic order can be written as\footnote{We adopt the same sign conventions as in 
\cite{GKSC}.}
\be 
\beta(\lambda) = \delta \lambda + \pi \tilde C \lambda^2\,.
\ee
The new fixed point is located at 
$$
\lambda^{*} = -\frac{\delta}{\pi \tilde C}\,,
$$
which is small if $\delta \ll 1$ and the scheme-dependent coefficient $\tilde C$ is fixed.
In practical applications, such as flows between neighbouring minimal models 
${\cal M}_m$ and ${\cal M}_{m-1}$ with $m\gg 1$, $\delta$ and $\tilde C$ both depend on a 
small parameter like $1/m$. We can trade this parameter for $\delta$ and thus $\tilde C = \tilde C(\delta)$.  
In such cases, one obtains a small $\lambda^*$ if the limiting value of $\tilde C$ is
non-vanishing in the $\delta \to 0$ limit.
In practice we can always choose a scheme, which we will call Wilsonian,  in which $\tilde C=C$ and 
this assumption is easy to check.  Thus if we perturb the system in the right direction we will 
flow to an infrared fixed point at $\lambda=\lambda^*$ which is nearby in the coupling 
space.\footnote{We can make the concept of a distance in theory space more precise by using the 
Zamolodchikov metric. In the scheme in which the metric is held fixed we find that the distance in 
the Zamolodchikov metric between the two fixed points is proportional to $\lambda^*$.}

We consider the perturbation defect associated to such a flow. Since $\phi$ has dimension close to 
two, the only relevant perturbation from the boundary point of view is the identity field. Thus, 
after the introduction of the bulk counter terms, only additional linear divergences can be 
present along the position of the defect. To get the conformal defect we therefore only need to 
subtract these additional linear divergences and arrive at the bulk fixed point $\lambda^*$. We 
calculate the $g$-factor of Affleck and Ludwig \cite{AL1} for the conformal perturbation defect at 
the leading and the next-to-leading orders. Our main result can be formulated as follows. In the 
Wilsonian RG scheme  the beta function up to cubic order has the form 
  \be
  \beta(\lambda)= \delta \lambda + \pi C \lambda^2 + \pi^2 D \lambda^3\,,
  \ee
where the coefficient $D$ is universal (scheme-independent) up to terms of the order 
${\cal O}(\delta)$. We find that the square of the  perturbation defect $g$-factor equals 
\be \label{main_result}
g^2_{\phi}= 1 +  \frac{\delta^2}{2  C^2} + \frac{\delta^3 D}{ C^4}   + {\cal O}\left(\delta^4 \right)\, . 
\ee
This formula is our main result. It is a universal formula of the type derived in \cite{CL}, 
\cite{Zamolodchikov} for the change of the central charge along the short flows. Another  formula of
similar type was derived  in \cite{AL2} for the change of the $g$-factor along short boundary 
flows.

The leading order correction in \eqref{main_result},
\be \label{g_leading}
g^2_{\phi}= 1 +  \frac{\delta^2}{2  C^2} +  {\cal O}\left(\delta^3 \right)\,,
\ee
is given by  (the limiting value of) the OPE coefficient $C$ in (\ref{OPE}). It is interesting to 
note that in a unitary theory the leading order correction is always positive, unlike in pure 
boundary flows \cite{AL1,AL2}. As is well known to condensed matter theorists the 
$g$-theorem of \cite{AL1,FK} can be violated in the bulk plus boundary RG flows (see 
{\it e.g.} \cite{Florens_Rosch}).

Specialising formula (\ref{main_result}) to the flows between 
the neighbouring  minimal models \cite{BPZ,Zamolodchikov}
we checked it against the $g$-factor of the RG domain wall 
candidate constructed in \cite{Gaiotto} and found the exact match in the leading and 
next-to-leading orders in the expansion parameter. This corroborates that the defect of \cite{Gaiotto} 
is the perturbation defect for $\phi_{1,3}$ flows in the minimal models. We hope that the 
perturbation defect picture will be more suggestive in understanding the relationship with the RG 
coefficients $b_i^{\;j}$.

The rest of the paper is organised as follows. In Section \ref{g} we discuss how to set up a 
perturbative expansion for the $g$-factor of the perturbation defect. In Section \ref{ren} we discuss the 
scheme independence and a convenient choice of renormalization scheme. In Sections \ref{quadratic} 
and \ref{cubic} we outline the calculation of the leading and next-to-leading order contributions. 
In Section \ref{G_defect} we discuss our result in relationship with the defect constructed in \cite{Gaiotto}. 
We conclude in Section \ref{conclusion} with some remarks. All hard computational details are moved 
to the appendices.

\section{Boundary entropy of perturbation defects} \label{g}
We consider the perturbation defect as a boundary condition in a tensor product of the undeformed 
${\rm CFT}_{\rm UV}$ and the deformed theory which we will drive to the IR fixed point. In order to
calculate the boundary entropy we follow \cite{AL1} and put this boundary condition on a cylinder of 
length $L$ and circumference $\beta$. From the cylinder partition function we can extract the 
boundary partition function as the finite piece in the $L\to \infty$ limit. At the IR fixed point the value 
of the boundary partition function gives the $g$-factor, and the boundary entropy is simply $\ln g$. 
If we unfold the defect, the cylinder becomes a torus in which half of the torus is perturbed. We can 
set up a perturbative expansion for the free energy of such a torus as  
 \be \label{torus_pert}
 \ln \frac{Z(L)}{Z_{\rm UV}(L)} = \langle e^{\delta S} \rangle_{\rm UV} \, , 
 \qquad \delta S = \lambda {\ell}^{-\delta} \int\limits_{T_{1/2}}\!\! d^2 w\, \phi(w) \,,
 \ee
 where we introduced a dimensionless coupling $\lambda$, $\ell$ is a renormalization length scale,  
 $T_{1/2}$ stands for the perturbed half torus, and $\langle . \rangle_{\rm UV}$ denotes the
 connected correlators evaluated in ${\rm CFT}_{\rm UV}$ on the torus.  For  generic small value of 
 $\delta$ all UV divergences in this perturbation series are power-like divergences which we can 
 unambiguously subtract, {\it e.g.} using analytic continuation in $\delta$. 

As we take the limit $L\to \infty$ the torus becomes very long and the leading contribution must 
come from a cylinder partition function with the ends capped by the vacuum of the UV theory 
$|0\rangle$ (see the picture below).\footnote{Strictly speaking we are assuming that no pathology 
like exponentially growing overlaps with excited states develops.}

\noindent \includegraphics{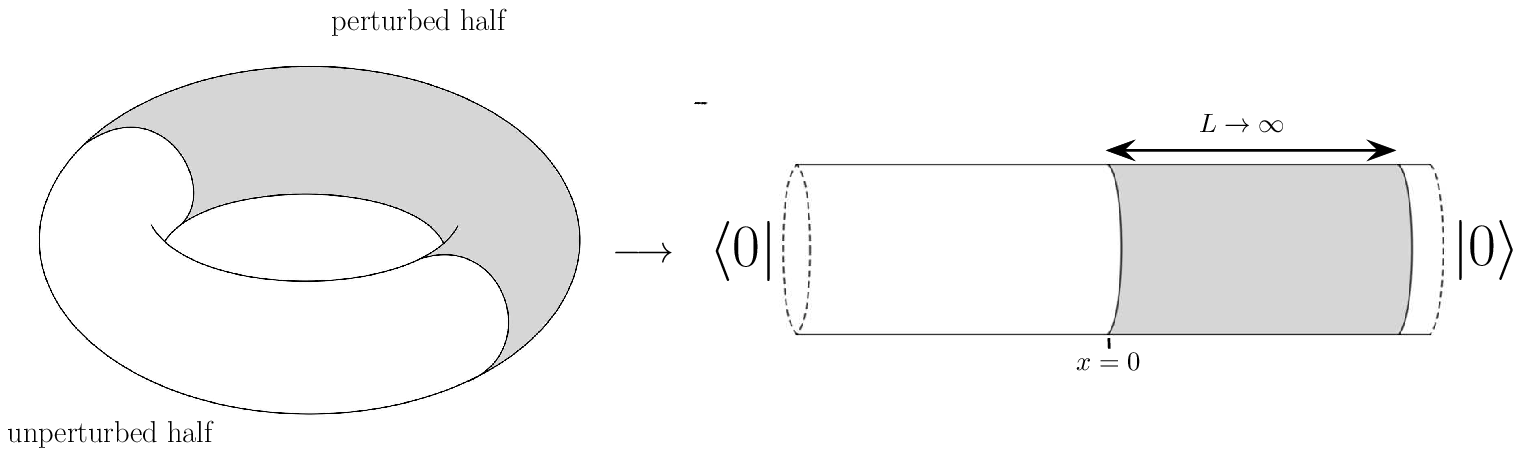}\\
Let $w=x+i\tau$ stand for a coordinate on an infinite cylinder. We obtain 
 \be\label{logZ}
 \ln Z = \left(\frac{L}{\beta}\right)  \frac{c_{\rm UV}\,\pi}{6} + \sum_{n=1}^{\infty} 
 (\lambda \ell^{-\delta})^{n} F_{n}   + \dots \,,
 \ee
where the ellipsis stand for the terms exponentially suppressed in $L$.  Here 
\be 
F_{n} = \frac{1}{n!} \int\limits_{0\le x_1\le L}\!\!d^2 w_1 \dots \int\limits_{0\le x_n\le L} %
\!\!d^2w_n\,  \langle \phi(w_1) \dots \phi(w_n)\rangle_{0}\,,
\ee
where $\langle .\rangle_0$ stands for the connected correlator in ${\rm CFT}_{\rm UV}$ at finite temperature $1/\beta$, and   $c_{\rm UV}$ is the central charge of the UV theory. 
We can easily separate the extensive part in $L$ in each of the integrals $F_{n}$. To that 
end we first use the permutation symmetry of the $n$-point function to fix the order along 
the cylinder axis:
\be 
F_n = \hspace{-0.5cm}  \int\limits_{0\le x_1\le x_2 \dots \le L} \hspace{-0.5cm}  d^2w_1d^2w_2 
\dots d^2w_n \, \langle \phi(w_1) \dots \phi(w_n)\rangle_{0} \,.
\ee
Next we introduce the variables 
  $$
  \xi_1 = w_2 - w_1\,   , \xi_2 =w_3 - w_1\, , \dots   \, , \xi_{n-1} = w_n-w_1\, . 
  $$
Due to translational and rotational invariance we have 
  $$
  \langle \phi(w_1) \dots \phi(w_n)\rangle_{0} = G_n(\xi_1, \dots \xi_{n-1}) \, . 
  $$ 
We can thus integrate out the $w_1$ variable explicitly, keeping the variables $\xi_i$ 
fixed. The range of integration for $w_1$ is $0\le x_1\, \le L-{\rm Re}\,  \xi_{n-1}$, 
$0\le \tau_1 \le \beta$.  We obtain $F_n = f_n + g_n$ with 
    \be
    f_n = \beta L \hspace{-1cm} \int\limits_{0\le {\rm Re} \xi_1 \le {\rm Re} \xi_2 \dots \le L} %
    \hspace{-1cm} d^2 \xi_1  \dots d^2\xi_{n-1} G_{n}(\xi_1, \dots , \xi_{n-1}) 
    \ee
and 
    \be \label{gn}
    g_n = - \beta \hspace{-1cm} \int\limits_{0\le {\rm Re} \xi_1 \le {\rm Re} \xi_2 \dots \le L} %
   \hspace{-1cm}  d^2 \xi_1 \dots d^2\xi_{n-1} \, ({\rm Re}\,  \xi_{n-1} )
   G_{n}(\xi_1, \dots , \xi_{n-1}) \, . 
    \ee
In the limit $L\to \infty$, the extensive contributions from the series with $f_n$ give the 
ground state energy correction, which at the IR fixed point is proportional to the change 
in the central charge. The perturbative integrals $f_2$ and $f_3$ were used in \cite{CL} 
to derive the difference of the central charge at leading order, 
\be
c_{\rm UV} - c_{\rm IR} = \frac{\delta^3}{C^2} + {\cal O}(\delta^4) \, . 
\ee
This formula was also derived in \cite{Zamolodchikov} by a different method. 

The non-extensive contributions $g_n$ at the IR fixed point sum up to $\ln g_{\phi}^2$ 
--- the boundary entropy of the perturbation defect. We are going to use the integrals 
$g_2$ and $g_3$ to calculate the first two terms in the $\delta$-expansion of $g^2_{\phi}$. 
This quantity is in some sense a natural companion of the central charge shift. We note that 
the finite temperature correlation functions at hand decay exponentially at $L\to \infty$,
so that there can be no extra contributions from the $f_n$ integrals to the finite $g^2_{\phi}$ 
piece. In the next section we discuss in detail the renormalization scheme (in)dependence 
of the calculation.

\section{Renormalization} \label{ren}
As mentioned in the introduction, for generic $\delta\ll 1$ we only have power divergences, 
which can be subtracted by analytic continuation in $\delta$. However, at $\delta =0$ the coupling 
becomes marginal and we get  logarithmic divergences. For finite $\delta$ they show up as poles 
in renormalised quantities  at $\delta=0$. To obtain a small $\delta$ expansion of any physical 
quantity we must subtract these poles. This defines a good  coordinate near the IR fixed point. 
We consider the beta function in a family of schemes in which the $\delta=0$ singularities are 
subtracted. We call such schemes Wilsonian.
The beta function has  the form 
\be\label{beta_gen}
\beta(\lambda) = \delta \lambda + \pi \tilde C \lambda^2 + \pi^2 D \lambda^3 + {\cal O}(\lambda^3) \,.
\ee
The coefficient $\tilde C$ is scheme dependent. To make the discussion concrete consider the 
short-distance contribution inside deformed correlators from two colliding operators $\phi$.  
Using the OPE (\ref{OPE}) we have
\be
\frac{\lambda^2}{2!} \ell^{-2\delta} \int_{0 \le |r| \le \epsilon} \!\! d^2 z \frac{C}{r^{\Delta} } \phi(0)  \,,
\ee
where $\epsilon$ is some short distance scale. Integrating we obtain 
\be
\frac{\pi C  }{\delta}  \left( \frac{\epsilon}{\ell}\right)^{\delta} (\ell^{-\delta} \phi(0) ) \, . 
\ee
A counterterm that subtracts the short-distance singularity is 
\be \label{tct}
S_{\rm ct} = - \lambda^2 \pi C t^{\delta}\frac{1}{\delta} \int \!\! d^2z\, \ell^{-\delta} \phi(z)\,,
\ee
where $t= \epsilon/ \ell$ is some arbitrary parameter. 
The corresponding beta function is 
\be
\beta(\lambda) = \delta \lambda + \pi C t^{\delta} \lambda^2 \,,
\ee
so that $\tilde C = \pi C t^{\delta}$. Note  that in $\delta \to 0$ limit the pole part of the subtraction 
is universal. Respectively, the value of   $\tilde C$ at $\delta =0$ is universal 
and given by the limiting value of the OPE coefficient $C$:
$$
\lim_{\delta\to 0} \tilde C= C_0=\lim_{\delta \rightarrow 0} C\, .
$$
We will assume throughout the discussion that $C_0\ne 0$.
 
More generally we can add a finite one-loop counterterm by hand to \eqref{tct},
provided it is non-singular at $\delta \to 0$. Explicitly, consider adding 
\be \label{ct_shift}
\Delta S_{\rm ct} = -\lambda^2 \pi f(\delta) \int \!\! d^2z\, \ell^{-\delta} \phi(z)\,,
\ee
where $f(\delta)$ is a function which is non-singular at $\delta=0$. The corresponding 
beta function undergoes a change in the quadratic coefficient,
\be
\tilde C \mapsto \tilde C + \delta  f(\delta)\, . 
\ee
 Introducing the expansion 
 \be
 \tilde C = C_0 + \tilde C_1 \delta + {\cal O}(\delta^2) 
 \ee
 we see that (\ref{ct_shift}) results in shifting 
 \be \label{C1_shift}
 \tilde C_1 \mapsto \tilde C_1 + f(0) \, .  
 \ee
Similarly to the above discussion of $\tilde C$ the value
$$
D_0 = \lim_{\delta \rightarrow 0} D
$$
is universal for any Wilsonian scheme.  The scheme dependent part 
of $D$ is of the order of $\delta$.  The infrared fixed point corresponding 
to (\ref{beta_gen}) is located at 
\be \label{lambda_IR}
\lambda^{*} = - \frac{\delta}{\pi \tilde C} - \frac{\delta^2D}{\pi \tilde C^3} + {\cal O}(\delta^3)\,.
\ee 
We see that at $\lambda=\lambda^*$ the scheme dependent contribution from $D$  
goes as $\delta^4$. We will be working to the order $\delta^3$ and therefore can neglect 
the scheme dependent part of $D$. \\

Let us now turn to the perturbative calculation of $g^2_{\phi}$ for the perturbation defect. 
On general grounds we expect $g_{\phi}$ to be scheme independent. It will 
be instrumental however to demonstrate this explicitly. To the order $\delta^3$ it suffices 
to consider only the quadratic and cubic terms:
 \be
 g_{\phi}^2 = 1 + (\lambda \ell^{-\delta})^2 g_{2}^{\rm ren} + (\lambda \ell^{-\delta})^3 g_{3}^{\rm ren} 
 + \dots \,,
  \ee
where $g_2^{\rm ren}$ and $g_3^{\rm ren}$ stand for the renormalised (subtracted) values 
of the corresponding integrals given in (\ref{gn}).\footnote{Notice that \eqref{logZ} actually 
computes the logarithm of $g$. However, to the order in perturbation in which we are working
the coefficients $g_n$ are the same.} As we will demonstrate in the next section,
\be
g_2^{\rm ren} = \left( \frac{\beta}{2\pi}\right)^{2\delta} I_{2}^{\rm ren}(\delta) \, , \qquad 
g_3^{\rm ren} = \left( \frac{\beta}{2\pi}\right)^{3\delta} I_{3}^{\rm ren}(\delta)\,,
\ee
where $I_2(\delta)^{\rm ren}$, $I_3^{\rm ren}(\delta)$ are subtracted integrals which  are numerical
functions, independent of any dimensionful  parameters.  The quantity $g_2^2$ contains only power 
divergences related to the identity field in the bulk and on the boundary, and its analytic continuation 
has no singularity at $\delta=0$.  
The cubic term $g_3$ contains a pole $\sim 1/\delta$ coming from two $\phi$-insertions colliding 
away from the boundary. This pole is to be subtracted by the same bulk counterterm used to renormalise 
the bulk theory. By naive power counting there can be no extra poles from collisions at the boundary of 
the integration region. Analyzing $g_3$ in Section \ref{cubic}, we will check  this as well as the correct 
factorisation of $g_3$ at the pole explicitly. 

It is convenient to absorb the factors of 
\be
s^{\delta}= \left(\frac{\beta}{2\pi \ell}\right)^{\delta}
\ee
into a rescaled coupling constant 
\be
\lambda \to \lambda s^{\delta}\, . 
\ee
To the order $\delta^3$ such a rescaling only shifts the $\tilde C_1$ coefficient in the beta function. 
After the rescaling we have 
\be \label{g_22}
g_{\phi}^2 = 1 + \lambda^2 I_2^{\rm ren}(\delta) + \lambda^3 I_3^{\rm ren}(\delta) + \dots \,.
\ee
Writing 
\be
I_2^{\rm ren}(\delta) = I_2^{(0)} + \delta I_{2}^{(1)} + {\cal O}(\delta^2)  \, , 
\qquad I_{3}^{\rm ren}(\delta) = I_3^{(0)} + {\cal O}(\delta) 
\ee
we can substitute (\ref{lambda_IR}) into (\ref{g_22}) and collect all terms up to the order $\delta^3$:
\be \label{g2_detailed}
g^{2}_{\phi} = 1 + \frac{\delta^2 I_{2}^{(0)}}{\pi^2 C_0^2} + \frac{\delta^3}{\pi^3 C_0^3} 
\Bigl[    I_2^{(1)}\pi C_0 +  \frac{2\pi D_0 I_{2}^{(0)}}{C_0} -(I_3^{(0)} + 2\pi \tilde C_1 I_2^{(0)}) 
\Bigr] + {\cal O}(\delta^4) \,.
\ee
We see that the only scheme dependent terms in this expression are $\tilde C_1$ and $I_3^{(0)}$.
It is easy to see from the definition (\ref{gn}) that the change in the bulk 1-loop counterterm 
(\ref{ct_shift}) will result in the shift 
\be
I_3^{(0)} \to I_3 -  2\pi f(0) I_{2}^{(0)}  \,.
\ee
Taking into account (\ref{C1_shift}) we therefore observe that the combination 
\be
I_3^{(0)} + 2\pi \tilde C_1 I_2^{(0)}
\ee
is scheme independent, and so is the expansion (\ref{g2_detailed}).

The scheme independence means we can use any convenient scheme to calculate 
$I_2^{\rm ren}$, $I_3^{\rm ten}$. In the following we will choose the scheme given by (\ref{tct}) 
with $t=1$, which we will call a Wilsonian minimal scheme. In this scheme the detailed 
representation of $g_{\phi}^2$ has the form (\ref{g2_detailed}) with $\tilde C_1 = C_1$,
where $C_1$ is the coefficient from the expansion 
\be
C=C_0 + C_1 \delta + {\cal O}(\delta^2) \, . 
\ee

\section{Quadratic order} \label{quadratic}
We start with the expression 
\be
g_2 = -\beta \int\limits_{0}^{\infty}\!\!dx \int\limits_{0}^{\beta}d\tau x 
\frac{(\pi/\beta)^{2\Delta}}{\left| \sinh\left(\frac{\pi(x+i\tau)}{\beta}\right)\right|^{2\Delta} } \, .
\ee
Mapping the infinite half-cylinder $x\ge 0$   onto the unit disc with coordinate 
$$
\eta = e^{-\frac{2\pi}{\beta} \xi} 
$$
we obtain 
\be
g_2 = \left(\frac{2\pi}{\beta}\right)^{2\delta} I_{2}(\delta)
\ee
with 
\be
I_2(\delta)  = 2\pi\!\! \int\limits_{|\eta|\le 1} \!\! d^2 \eta\, 
\frac{\ln |\eta|}{|\eta|^{\delta}|1-\eta|^{2(2-\delta)} }\, . 
\ee
This integral converges for $\frac{1}{2}<\delta <2$.
We show in Appendix A that the analytically continued $I_2(\delta)$ has the expansion
\be\label{I2}
I_2(\delta)^{\rm ren} = \frac{\pi^2}{2} + \frac{3\pi^2}{2}\delta + {\cal O}(\delta^2) \,,
\ee 
and hence 
\be \label{i22}
I_2^{(0)}=\frac{\pi^2}{2}\, , \qquad I_2^{(1)} = \frac{3\pi^2}{2}\, . 
\ee
Substituting the value of $I_2^{(0)}$ into the order $\delta^2 $ terms in (\ref{g2_detailed}) we 
obtain the leading order formula (\ref{g_leading}). It is also easy to obtain the value of $I_2^{(0)}$ 
from directly working at $\delta=0$ with a position-space cutoff. The integral has  a linear divergence 
coming from the region near the boundary of the infinite half-cylinder (or the disc).

\section{Cubic order} \label{cubic}
In the cubic order we are dealing with a double integral over an infinite half cylinder
\be
g_3 =-\beta C\left(\frac{\pi}{\beta}\right)^{3\Delta}  \iint\limits_{0\le{\rm Re}\xi_1 
\le {\rm Re}\xi_2 \le \infty} d^2\xi_1 d^2\xi_2  \, ({\rm Re}\, \xi_2) G(\xi_1,\xi_2) \,,
\ee
where 
\be
G(\xi_1,\xi_2)=\left|\sinh\left(\frac{\pi\xi_1}{\beta} \right)\sinh\left(\frac{\pi\xi_2}{\beta} \right)
\sinh\left(\frac{\pi(\xi_1-\xi_2)}{\beta}  \right)\right|^{-\Delta} \, . 
\ee
Using the variables 
\be
\nu_1 = e^{-\frac{2\pi}{\beta}}\, , \qquad \nu_2 = e^{-\frac{2\pi}{\beta}(\xi_2 - \xi_1)} 
\ee
we obtain 
\be
g_3 = \left( \frac{\beta}{2\pi} \right)^{3\delta} I_{3}(\delta) \,, 
\qquad I_{3}(\delta) = C i_{3}(\delta)
\ee
with 
\be\label{i3_disc}
i_3(\delta) = 4\pi \iint\limits_{|\nu_1|\le 1, |\nu_2|\le 1} d^2\nu_1d^2\nu_2\, 
\frac{|\nu_1\nu_2|^{-\delta}\ln|\nu_1|}{|(1-\nu_1)(1-\nu_2)(1-\nu_1\nu_2)|^{2-\delta}}\, . 
\ee
As we show in Appendix B, this integral converges for $1/3<\delta<2$.\footnote{There is an upper bound
because we removed the IR cutoff. The lower bound can be obtained from the OPE estimates.}
The analytically continued function has the following expansion near $\delta=0$:
\be\label{i3_1}
i_{3}(\delta) = \frac{\pi^3}{\delta} + \frac{9}{2}\pi^3 + {\cal O}(\delta) \, .
\ee
The calculation leading to (\ref{i3_1}) is quite long and tedious and is presented in 
Appendix~B.%

From (\ref{i3_1})  we  observe that the pole in $g_3$ can be written as
\be
g_3 \sim 2\pi C_0  I_{2}^{(0)} \frac{1}{\delta} + \dots \,,
\ee 
and therefore it is indeed subtracted by the bulk counterterm (\ref{tct}). This proves that, as 
expected by power counting, no additional  renormalization is needed for $g_{\phi}$ apart 
from the analytic continuation that gets rid of linear divergences. In the  Wilsonian minimal scheme 
(described at the end of section \ref{ren}) we 
subtract $CI_{2}^{\rm ren}(\delta)$ at the pole. Using  (\ref{I2}) we rewrite (\ref{i3_1}) as 
\be
i_3(\delta) = \frac{\pi I_{2}^{\rm ren}(\delta)}{\delta} + \frac{3}{2}\pi^3 + {\cal O}(\delta) \, . 
\ee
Hence in this scheme 
\be
I_{3}^{(0)} = \frac{3}{2}\pi^3 C_0\, . 
\ee
Substituting this along with (\ref{i22}) into (\ref{g2_detailed}) we obtain 
\be
g^2_{\phi} = 1 + \frac{\delta^2}{2C_0^2} + \frac{\delta^3}{C_0^4}( D_0 - C_0 C_1) 
+ {\cal O}(\delta^4)   
\ee
or, more compactly, 
\be
g_{\phi}^{2} = 1  +  \frac{\delta^2}{2  C^2} + \frac{\delta^3 D}{ C^4}   + {\cal O}\left(\delta^4 \right)\,,
\ee 
which is our main result.

\section{Gaiotto's defect} \label{G_defect}
We would like to apply our formula (\ref{main_result}) to flows between neighbouring 
A-series unitary  minimal models ${\cal M}_m$ which have the central charges 
\be
c_m=1-\frac{6}{m(m+1)}\,,\qquad m=3,\,4,\ldots\,.
\ee
The flow is triggered by the $\phi_{1,3}$ operator \cite{Zamolodchikov}: 
\be
{\cal M}_{m} \stackrel{\phi_{1,3}}{\longrightarrow} {\cal M}_{m-1}  \, . 
\ee
For large values of $m$ the flow can be studied perturbatively using $1/m$ as the 
expansion parameter. 
Alternatively one can use any other function of $m$ which is decreasing as $m\to \infty$. 
We will assume, for now without specifying, that we have a small parameter $1/k$.
Let 
\be
\delta = \frac{\delta_1}{k} + \frac{\delta_2}{k^2} + {\cal O}\left( \frac{1}{k^3}\right) 
\ee 
where $\delta_1$, $\delta_2$ are some numbers. Furthermore let 
\be
 C = C_0 +  C_1 \frac{\delta_1}{k} + {\cal O}\left( \frac{1}{k^3}\right) \, , 
 \qquad D = D_0 + {\cal O}\left( \frac{1}{k}\right) 
\ee
Formula (\ref{main_result}) can be rewritten as 
\be \label{g2_det_k}
g^2_{\phi}= 1 + \frac{1}{k^2}\left(\frac{\delta_1^2}{2C_0^2}\right) + \frac{1}{k^3}
\left( \frac{\delta_1}{C_0^4}  \right)[ \delta_2 C_0^2 + \delta_1^2(D_0 - C_0C_1)] 
+  {\cal O}\left(\frac{1}{k^4} \right)\, .
\ee
This formula can be applied to any short flows with a small parameter.
Specialising it to the minimal models we choose $k=m+2$ as in \cite{Gaiotto}. We have (see  \cite{DF, Zamolodchikov}) 
\be
\delta_1 =2\, , \qquad  \delta_2=-12 \,  , 
\ee
\be
C_0 = \frac{4}{\sqrt{3}} \, , \qquad C_1 =-\sqrt{3} \, . 
\ee
Furthermore,
\be
D_0 = -\frac{8}{3}\, . 
\ee
The last result can be taken from \cite{Zamolodchikov} (see also 
\cite{Lassig,Constantinescu-Flume} for detailed calculations). 
Substituting these values into (\ref{g2_det_k}) we obtain 
\be\label{min_pred}
g^2_{(1,3)} = 1 + \frac{3}{2k^2} - \frac{6}{k^3} + {\cal O}\left(\frac{1}{k^4}\right) \, .
\ee
The $g$-factor for the defect constructed by Gaiotto in \cite{Gaiotto} is
\bea\label{gG}
&& g^{2}_{\rm Gaiotto} = \frac{k+2}{\sqrt{(k+1)(k+3)}}\frac{\sin\left(\frac{\pi}{k+1}\right)
\sin\left(\frac{\pi}{k+3}\right)}{\sin^2\left(\frac{\pi}{k+2}\right)}\nonumber \\
&& = 1 + \frac{3}{2k^2} -  \frac{6}{k^3} + \left( -\pi^2 + \frac{159}{8} \right)
\frac{1}{k^4} + \dots 
\eea
We observe the precise match between (\ref{min_pred}) and the first two terms 
in the expansion (\ref{gG}).


\section{Concluding remarks}\label{conclusion}
As the calculation in the preceding section shows, the defect constructed in \cite{Gaiotto} 
algebraically has a good chance of being a conformal perturbation defect accompanying 
the short RG flows in the minimal models. Our general formula (\ref{main_result}) can be 
used to check candidate defects for other short flows, {\it e.g.} for the coset flows considered 
in \cite{CSS}. 
 
It would be interesting to pursue the construction of perturbation defects further. In particular, 
instead of the perturbed cylinder amplitude capped by the vacuum states we could consider 
the cylinder amplitudes between excited  states in the UV theory. It seems plausible to us 
that such overlaps contain matrix elements for $b\cdot b^{T}$ where $b=(b_i^{\;j})$ is defined 
in (\ref{bij}). However we do not understand the right RG scheme in which one should 
calculate the $b_i^{\;j}$. 

At  $m=\infty$  the coefficients $b_i^{\;j}$ for the flows 
between the minimal models are 
scheme independent, and there is an impressive matching \cite{Gaiotto}, \cite{Poghossian^2} with the 
corresponding 1-point functions for the RG defect proposed by Gaiotto. It is 
not clear to us however that the $m=\infty$ matching should be taken as a smoking gun of 
the RG defect. At $m=\infty$ the two neighbouring minimal models are isomorphic and the 
coefficients $b_i^{;j}$ establish a particular isomorphism associated with the perturbing 
operator $\psi_{1,3}$ (which at $m=\infty$ becomes marginal). It should be noted that for 
finite $m$ there is currently no understanding of how to match the overlaps of \cite{Gaiotto} 
with the RG based calculations \cite{Poghossian}. We hope that the general picture of an 
RG defect as a perturbation defect may help to undertand the possible relationship between 
these quantities.

It would be very interesting to understand the 
intrinsic field theoretic meaning of the number $g_{\phi}$ that in principle is associated with 
every bulk RG flow.\footnote{One could speculate that the logarithm of $g_{\phi}$ may be 
proportional to the length of the RG trajectory in Zamolodchikov's metric. But for the minimal 
model short flows the length of the trajectory goes as $1/m$ while the leading term in 
$g_{\phi}$ goes as $1/m^2$, so any relationship between the distance and $g_{\phi}$, if it 
exists, must be less straightforward.} Hopefully we will be able to clarify this question in future 
work.


\setcounter{equation}{0}

\begin{center}
{\bf Acknowledgements}
\end{center}

We thank Charles Melby-Thompson for useful discussions. The work of CSC was supported by the World 
Premier International Research Center Initiative (WPI), MEXT, Japan, and he is grateful for the hospitality 
at Heriot-Watt University, Edinburgh, UK. 
The work of AK was supported in part by STFC grant 
ST/J000310/1 ``High energy physics at the Tait Institute''.


\appendix
\renewcommand{\theequation}{\Alph{section}.\arabic{equation}}
\setcounter{equation}{0}

\section{The integral $I_{2}$}
Before we present all technical details pertaining to evaluating the integrals $I_2(\delta)$ 
and $i_3(\delta)$ we would like to make a general remark about integrating various 
power series expansions. The integrals we will be computing will all split into subintegrals 
each calculated over a half-disc  
\be
{\mathbb D}^{+}= \{ z \in {\mathbb C} : {\rm Im}\, z \ge 0\, , |z|\le 1\} \, .
\ee
The integrands will typically contain functions which can be expanded about the centre of 
${\mathbb D}^{+}$ and whose radius of convergence may be equal to 1. For all such situations 
we find that if one first does the angular integration (keeping the region of integration to be a half 
disc of a radius a bit smaller than 1) then the remaining series in the powers of the radius has a 
radius of convergence greater than 1 and thus can be integrated term by term over ${\mathbb D}^{+}$.\\

For $I_2$ our starting point is formula 
\be
I_2(\delta) = 2\pi \int\limits_{|\eta|\le 1}\!\!d^2\eta \, \frac{\ln |\eta|}{|\eta|^{2-\Delta}|1-\eta|^{2\Delta}}
\ee
Mapping this to the half-plane with coordinate 
\be
z=\frac{i(1-\eta)}{1+\eta}
\ee
we obtain 
\be
I_{2}(\delta) = \pi 2^{2\delta-1}\left(\frac{\beta}{\pi}\right)^{2\delta}\displaystyle{\int_{\im z\geq 0}}
d^2z\,\ln\left|\frac{1+iz}{1-iz}\right|\,\frac{|1+z^2|^{\Delta-2}}{|z|^{2\Delta}}\,.
\ee
Splitting the half-plane into the regions $|z|>1$ and $|z|<1$ we rewrite the $I_2$ as an integral 
over the upper half unit disc:
\be
I_2(\delta) = \pi 2^{2\delta-1} \displaystyle{\int_{\mathbb{D}^+}} d^2z\,\ln\left|\frac{1+iz}{1-iz}\right|\,
|1+z^2|^{\Delta-2}\left(\frac{1}{|z|^{2\Delta}}+1\right)\, .
\ee
We will find a $\delta \to 0$ expansion for a more general integral 
\be \label{Iint}
I(x,\delta) = 2^{2\delta}\int_{\mathbb{D^+}}d^2z\,\ln\left|\frac{1+iz}{1-iz}\right|\,|1+z^2|^{-\delta}\,
\left(1+|z|^{-4+x\delta}\right)
\ee
where $x\ge 2$. We have $I_2(\delta)=\frac{\pi}{2}I(2,\delta)$. 

The integral (\ref{Iint}) converges for $1/2<\delta < 2$. It is not hard to see that the analytic 
continuation from this region commutes with expanding the factor $|1+z^2|^{-\delta}$ for small 
$\delta$. We can use 
\be\label{first}
I(x,\delta)= \int\limits_{\mathbb{D^+}}\!\!d^2z\,\ln\left|\frac{1+iz}{1-iz}\right|\left(1+|z|^{-4+x\delta}\right)
-\delta\!\!\int\limits_{\mathbb{D}^+}\!\!d^2z\,\ln\left|\frac{1+iz}{1-iz}\right|\,
\ln|1+z^2|\,\left(1+|z|^{-4}\right)+\mathcal{O}(\delta^2)
\ee
For the first integral in (\ref{first}) we obtain using the expansion (\ref{exp}) and analytic continuation 
\bea \label{polovina}
&& \int_{\mathbb{D^+}}d^2z\,\ln\left|\frac{1+iz}{1-iz}\right|\left(1+|z|^{-4+x\delta}\right)=
\sum_{p=0}^\infty\frac{4(-1)^{p+1}}{(2p+1)^2}\left(\frac{1}{2p+3}+\frac{1}{2p-1+x\delta}\right) \nonumber \\
&&= \sum_{p=0}^\infty\frac{8(-1)^{p+1}}{(2p-1)(2p+1)(2p+3)}+
\sum_{p=0}^\infty\frac{4(-1)^{p}\,x\,\delta}{(2p+1)^2(2p-1)^2}+\mathcal{O}(\delta^2)\nonumber \\
&& =\pi + \frac{x}{2}(\pi+4)\delta+\mathcal{O}(\delta^2)\, .
\eea
For the second integral in (\ref{first}) we compute, using the power series expansions for the logarithms, 
\bea
&& I_{(2)} = \int\limits_{\mathbb{D}^+}\!\!d^2z\,\ln\left|\frac{1+iz}{1-iz}\right|\,
\ln|1+z^2|\,\left(1+|z|^{-4}\right) \\
&& \sum_{p,k=0}^\infty\frac{4(-1)^{p+k}}%
{(2k+1-2p)(k+1)(2k+3+2p)}\left(\frac{1}{(2k+5+2p)}+\frac{1}{(2k+1+2p)}\right) \nonumber 
\eea
We next change the summation variables and rewrite the last series so that the denominators are  
all positive:
\begin{align}
I_{(2)}=&\;\sum_{p,k=0}^\infty \frac{4(-1)^k}{(k+p+1)(2k+1)(2k+3+4p)}
\left(\frac{1}{2k+5+4p}+\frac{1}{2k+1+4p}\right)\nonumber\\
&\quad+\sum_{p,k=0}^\infty \frac{4(-1)^k}{(p+1)(2k+1)(2k+5+4p)}
\left(\frac{1}{2k+7+4p}+\frac{1}{2k+3+4p}\right)\,.
\end{align}
Next we repeatedly decompose the fractions so that eventually all denominators of each summand 
contain only one factor that involves both summation variables. We arrive at 
\begin{align}
&& I_{(2)}=\sum_{p,k=0}^\infty \frac{2(-1)^k}{(2k+1)}\bigg[
\frac{2}{(2p+3)(2k+5+4p)}-\frac{1}{(2p+3)(k+p+1)}+\frac{1}{(2p-1)(k+p+1)}\nonumber \\
&&-\frac{2}{(2p-1)(2k+1+4p)}-\frac{1}{(p+1)(2k+7+4p)}+\frac{1}{(p+1)(2k+3+4p)}
\bigg]\,. 
\end{align}
In each summand we write the factor containing both summation indices
as an integral
\begin{equation}
\frac{1}{n_1 k+n_2+n_3 p}=\int_0^1r^{n_1k+n_2+n_3p-1}\,dr\,,
\end{equation}
and then factor the summations under the integral. This gives us 
\begin{align}\label{almostL}
I_{(2)}=2\int_0^1dr\bigg[&
\sum_{k=0}^\infty\frac{(-1)^kr^k}{2k+1}\left(\frac{1}{r}-1+\left(r-\frac{1}{r}\right)
\sum_{p=0}^\infty\frac{r^p}{2p+1}
\right)\nonumber\\
&+\sum_{k=0}^\infty\frac{(-1)^kr^{2k}}{2k+1}\left(\left(1-r^4\right)\sum_{p=0}^\infty\frac{2r^{4p}}{2p+1}
+\left(r^2-r^6\right)\sum_{p=0}^\infty\frac{r^{4p}}{p+1}\right)
\bigg]\,.
\end{align}
Using the standard series expressions
\begin{equation}
\sum_{k=0}^\infty\frac{(-1)^kx^{2k}}{2k+1}=\frac{{\rm arctan}\,x}{x}\,,\quad
\sum_{p=0}^\infty\frac{x^{2p}}{2p+1}=\frac{{\rm atanh}\,x}{x}\,,\quad
\sum_{p=0}^\infty\frac{x^p}{p+1}=-\frac{\ln(1-x)}{x}
\end{equation}
we convert \eqref{almostL} into an integral of elementary functions.
 Integration  leads us to 
\begin{equation}\label{I2ndanswer}
I_{(2)}=4-2\pi+2\pi\ln 2 \; .
\end{equation}
Combining this with (\ref{polovina}) we finally obtain 
\begin{equation}\label{Iintegralanswer}
I(x,\delta)=\pi+\left[\frac{x}{2}(\pi+4)-4+2\pi-2\pi\ln 2\right]\,\delta+\mathcal{O}(\delta^2)\, ,
\end{equation}
and 
\be
I_2(\delta) = \frac{\pi}{2}I(2,\delta) = \frac{\pi^2}{2} + \frac{3\pi^2}{2}\delta + {\cal O}(\delta^2) \, .
\ee

\section{The integral $i_3$}
\setcounter{equation}{0}
\subsection{Splitting the region of integration}

In this appendix we analyse the integral $i_3$ given in (\ref{i3_disc}). We map 
the disc onto the upper half plane using the coordinates 
\be
z_i = i\frac{1-\nu_i}{1+\nu_i}\, , \enspace i=1,2. 
\ee
 We obtain 
 \be \label{i3_disc2}
 i_3 = \pi 2^{3\delta}\!\! \iint\limits_{z_{1},z_2 \in {\mathbb H}} d^2z_1d^2z_2\, 
 \ln\left|\frac{1+iz_1}{1-iz_1}  \right| 
 \frac{|1+z_1^2|^{-\delta}|1+z_2^2|^{-\delta}}{|z_1 z_2 (z_1 + z_2)|^{2-\delta}}
\ee
where ${\mathbb H}$ is the upper half plane ${\rm Im}\, z_i \ge 0$. 
Note that  standard methods, 
such as {\it e.g.} Feynman parameters, are of no use in   analysing  (\ref{i3_disc2}) because  most of the usual 
symmetries are broken, so we  resort to brute force.
To have control over divergences in various limits we split the region of integration into 
the following six regions
\bea
&& {\cal D}_{--}^{<}=\{ z_1,z_2 \in {\mathbb H}, |z_1|\le |z_2|\le 1\}\, , \quad 
{\cal D}_{--}^{>}=\{ z_1,z_2 \in {\mathbb H}, |z_2|\le |z_1|\le 1\}\nonumber \\
&& {\cal D}_{++}^{<}=\{ z_1,z_2 \in {\mathbb H}, 1\le |z_1|\le |z_2|\}\, , \quad  
{\cal D}_{++}^{>}=\{ z_1,z_2 \in {\mathbb H}, 1\le |z_2|\le |z_1|\}  \nonumber\\
&& {\cal D}_{+-}=\{ z_1,z_2 \in {\mathbb H}, |z_1|\ge 1\, , |z_2|\le 1 \}\, , \quad 
{\cal D}_{-+}=\{ z_1,z_2 \in {\mathbb H}, |z_2|\ge 1\, , |z_1|\le 1 \} \, . \nonumber \\
&&
\eea
We will denote the corresponding subintegrals as $I_{++}^{<}$ {\it etc}. 
Our final goal is to extract a pole and a finite part in the Laurent expansion near $\delta=0$. 
We observe that the integrals $I_{++}^{>}$, $I_{++}^{<}$, and $I_{-+}$ are finite in the $\delta\to 0$ 
limit. The $1/\delta$ pole comes from the regions ${\cal D}_{+-}$ and ${\cal D}_{--}^{>}$. The 
integrals $I_{--}^{>}$ and $I_{--}^{<}$ also contain power divergences in position space which are 
treated by analytic continuation in $\delta$. 

 \subsection{ The pole}
We first analyse the integrals $I_{+-}$ and $I_{--}^{>}$ and split each of them into a pole and a 
finite part at $\delta=0$:
\bea
I_{+-}(\delta) &&= \frac{I_{+-}^{p}}{\delta} + I_{+-}^{f} + {\cal O}(\delta) \, , \\
I_{--}^{>}(\delta) &&= \frac{I_{--}^{>,p}}{\delta} + I_{--}^{>,f} + {\cal O}(\delta) \, .
\eea
We start by looking at $I_{+-}$. The $z_2$-dependent factor in the integrand can be represented as 
\be\label{3terms}
\frac{|1+z_2^2|^{-\delta}}{|z_2(z_1+z_2)|^{\Delta}} = \frac{1}{|z_1z_2|^{\Delta}} + 
\frac{|1+ z_2^2|^{-\delta} - 1}{|z_1z_2|^{\Delta}} 
 + \frac{|1+z_2^2|^{-\delta}}{|z_2|^{\Delta}} \left( \frac{1}{|z_1 + z_2|^{\Delta}} - 
 \frac{1}{|z_1|^{\Delta}} \right)\, .
\ee
The pole comes from integrating the first term on the right hand side. Changing the variable to 
$u_1 = 1/\bar z_1$ we obtain  
\be
 \pi 2^{3\delta} \int\limits_{{\mathbb D}^{+}}d^2u_1 \, \ln\left| \frac{1 + iu_1}{1-iu_1}\right| 
 |1+u_1^2|^{-\delta} \int\limits_{{\mathbb D}^{+}}\frac{d^2 z_2}{|z_2|^{\Delta}}
 = \frac{I_{+-}^{p}}{\delta} + I_{++}^{f1} + {\cal O}(\delta) 
\ee
with
\bea \label{pole1}
I_{+-}^{p} = \pi^2 \int\limits_{{\mathbb D}^{+}}d^2u_1 \, \ln\left| \frac{1 + iu_1}{1-iu_1}\right| &&
 = -4\pi^2 \sum_{k=0}^{\infty} \frac{(-1)^k}{(2k+1)^2(2k+3)} = \nonumber \\
&& -2\pi^2( G -\frac{\pi}{4} + \frac{1}{2}) 
\eea
where $G$ is Catalan's constant
\be
G=\sum_{n=0}^\infty\frac{(-1)^n}{(2n+1)^2}\,.
\ee
The finite piece $I_{+-}^{f1}$ can be written as 
\be
I_{+-}^{f1} = 3\ln 2 I_{+-}^{p} + \lim_{\delta\to 0}\frac{1}{\delta}\Bigl[ \pi^2   
\int\limits_{{\mathbb D}^{+}}d^2u_1 \, \ln\left| \frac{1 + iu_1}{1-iu_1}\right| |1+u_1^2|^{-\delta} - I_{+-}^{p}\Bigr] \, .
\ee
An additional finite piece comes from the third term on the right hand side of (\ref{3terms})
\be
I_{+-}^{f2} = \pi  \int\limits_{{\mathbb D}^{+}}d^2u_1 \, \ln\left| \frac{1 + iu_1}{1-iu_1}\right| 
\int\limits_{{\mathbb D}^{+}}\frac{d^2z_2}{|z_2|^2}\left(\frac{1}{|1+\bar u_1 z_2|^2} - 1\right)\, .
\ee
The second term in (\ref{3terms}) will give no contribution at $\delta \to 0$ because the numerator 
can be expanded as 
\be
|1+z_2^2|^{-\delta} - 1 = -\delta [\ln(1 + w_2^2) + \ln(1+ \bar w_2^2)] + {\cal O}(\delta^2) 
\ee
and the leading term vanishes upon angular integration as all even powers integrate to zero on a 
half-disc. Hence
\be
I_{+-}^{f} = I_{+-}^{f1} + I_{+-}^{f2} \, .
\ee

We now turn to $I_{--}^{>}$. Changing the variable of integration from $z_2$ to $u_2 = \frac{z_2}{|z_1|}$ 
we split the integral over $u_2$ as 
\begin{align}\label{D-->decomposition}
\int\limits_{\mathbb{D}^+}d^2u_2\,\frac{|1+u_2^2\,|z_1|^2|^{\Delta-2}}{|u_2(e^{i\phi_1}+u_2)|^\Delta}&\,=\,
\int\limits_{\mathbb{D}^+}\frac{d^2u_2}{|u_2|^\Delta}
+ \int\limits_{\mathbb{D}^+}d^2u_2\,\frac{|1+u_2^2\,|z_1|^2|^{\Delta-2}-1}{|u_2|^\Delta}\\[5pt]
&\qquad+\int\limits_{\mathbb{D}^+}d^2u_2\,\frac{|1+u_2^2\,|z_1|^2|^{\Delta-2}}{|u_2|^\Delta}
\left(\frac{1}{|e^{i\phi_1}+u_2|^\Delta}-1\right)\,,\nonumber
\end{align}
where $\phi_1$ is the argument of $z_1$. The first term on the right hand side of (\ref{D-->decomposition}) 
contains a pole 
\be \label{I-->p}
I_{--}^{>,p} = \pi^2 {\lim_{\delta \to 0}}^{a} \int_{\mathbb{D}^+}
\frac{d^2z_1}{|z_1|^{3\Delta-2}}\,\ln\left|\frac{1+iz_1}{1-iz_1}\right|\,
\ee
 and a contribution to the finite part 
\be \label{I-->f1}
I_{--}^{>,f1}=3 \ln 2 I_{--}^{>,p}+  {\lim_{\delta\rightarrow0}}^a \, \frac{1}{\delta} \left[
\pi^2\int_{\mathbb{D}^+}\frac{d^2z_1}{|z_1|^{4-3\delta}}
\ln\left|\frac{1+iz_1}{1-iz_1}\right| |1+z_1^2|^{-\delta}\;-\;I_{--}^{>,p}
\right]\,.
\ee 
The integrals in (\ref{I-->p}) and (\ref{I-->f1}) should be analytically continued to $\delta=0$ from the region $1/3<\delta$. 
We denote such analytic continuations using the $\displaystyle{{\lim_{\delta\to 0}}^a}$ symbol. 

The second term in (\ref{D-->decomposition}) does not contribute to the finite part by virtue of angular integration. 
The third term gives a finite contribution after analytic continuation to $\delta=0$ from the region $1/3<\delta$:
\begin{equation}\label{I3-->f2}
I_{--}^{>,f2}={\lim_{\delta\to 0}}^{a}\, \pi\int_{\mathbb{D}^+}
\frac{d^2z_1}{|z_1|^{3\Delta-2}}\ln\left|\frac{1+iz_1}{1-iz_1}\right|
\int_{\mathbb{D}^+}\frac{d^2u_2}{|u_2|^2}\left(\frac{1}{|e^{i\phi_1}+u_2|^2}-1\right)\,,
\end{equation}
We evaluate the residue at the pole using (\ref{exp}),
\be\label{pole2}
I_{--}^{>,p}=-\frac{4\pi^2}{\delta}\sum_{k=0}^{\infty}\frac{(-1)^k}{(2k+1)^2(2k-1)}
=2\pi^2\left(G+\frac{\pi}{4}+\frac{1}{2}\right)\,.
\ee
Adding together (\ref{pole1}) and (\ref{pole2}) we obtain the complete residue 
\be\label{I3p}
I_{3}^{p} = I_{+-}^{p} + I_{--}^{>,p} = \pi^3 \,,
\ee
so that 
\be
i_{3}(\delta)= \frac{\pi^3}{\delta} + I_{3}^f + {\cal O}(\delta) \,,
\ee
where the finite part $I_3^f$ receives 8 contributions which are summarised in the following list:

\begin{table}[h!]
\begin{displaymath}
\!\!\!\!\!\!\!\!
\begin{array}{r@{\,=\,}lr}
I_{-+}&\displaystyle \pi\int_{\mathbb{D}^+}\frac{d^2z_1}{|z_1|^2}
\ln\left|\frac{1+iz_1}{1-iz_1}\right|
\int_{\mathbb{D}^+}\frac{d^2u_2}{|1+u_2\bar{z}_1|^2}\,,&\\[10pt]
I_{++}^<&\displaystyle \pi \int_{\mathbb{D}^+}d^2u_1
\ln\left|\frac{1+iu_1}{1-iu_1}\right|\int_{\mathbb{D}^+}
\frac{d^2v_2}{|e^{i\phi_1}+v_2|^2}&(u_1=|u_1|e^{i\phi_1})\,,\\[10pt]
I_{++}^>&\displaystyle \pi \int_{\mathbb{D}^+}d^2v_1\int_{\mathbb{D}^+}d^2u_2
 \ln\left|\frac{1+i|u_2|v_1}{1-i|u_2|v_1}\right|\frac{1}{|e^{i\phi_2}+v_1|^2}
 &(u_2=|u_2|e^{i\phi_2})\,,\\[10pt]
 I_{--}^<&\displaystyle \pi \int_{\mathbb{D}^+}\frac{d^2u_1}{|u_1|^2}
\int_{\mathbb{D}^+}\frac{d^2z_2}{|z_2|^{3\Delta-2}}
\ln\left|\frac{1+iu_1|z_2|}{1-iu_1|z_2|}\right|\frac{1}{|u_1+e^{i\phi_2}|^2}&
(z_2=|z_2|e^{i\phi_2})\,,\\[10pt]
I_{+-}^{f1}&\multicolumn{2}{l}{\displaystyle 3\ln2\,I_{+-}^p +
{\lim_{\delta\rightarrow0}} \frac{1}{\delta} \left[\pi^2\int_{\mathbb{D}^+}d^2u_1
\ln\left|\frac{1+iu_1}{1-iu_1}\right| |1+u_1^2|^{-\delta}\;-\;(I_3)_{+-}^p
\right]\,,}\\[10pt]
I_{+-}^{f2}&\displaystyle 6\pi\,
\int_{\mathbb{D}^+}d^2u_1\,
\ln\left|\frac{1+iu_1}{1-iu_1}\right| \int_{\mathbb{D}^+}\frac{d^2z_2}{|z_2|^2}
\left(\frac{1}{|1+\bar{u}_1z_2|^2}-1\right)\,,\\[10pt]
I_{--}^{>,f1}&\multicolumn{2}{l}{\displaystyle 3\ln2\,I_{--}^{>,p}
+{\lim_{\delta\rightarrow0}}^a\, \left[
\frac{\pi^2}{\delta}\int_{\mathbb{D}^+}\frac{d^2z_1}{|z_1|^{4-3\delta}}
\ln\left|\frac{1+iz_1}{1-iz_1}\right| |1+z_1^2|^{-\delta}\;-\;(I_3)_{--}^{>,p}
\right]\,,}\\[10pt]
I_{--}^{>,f2}&\displaystyle {\lim_{\delta\to 0}}^a\, \displaystyle \pi\int_{\mathbb{D}^+}
\frac{d^2z_1}{|z_1|^{3\Delta-2}}\ln\left|\frac{1+iz_1}{1-iz_1}\right|
\int_{\mathbb{D}^+}\frac{d^2u_2}{|u_2|^2}\left(\frac{1}{|e^{i\phi_1}+u_2|^2}-1\right)
&(z_1=|z_1|e^{i\phi_1})\,.
\end{array}
\end{displaymath}
\caption{The set of  integrals that contribute to $I_3^{f}$.}\label{allfiniteintegrals}
\end{table}
%
The following sections contain details of calculations of these 8 integrals.


\subsection{Computation of $I_{-+}$}

In $I_{-+}$ we find  the integral (\ref{J1}).
Using (\ref{J2}) and (\ref{exp}) we obtain 
\begin{equation}\label{I3pmintermediate}
I_{-+}=12\pi^2\sum_{p,k=0}^\infty\frac{(-1)^p}{(2p+1)^2}
\left(\frac{1}{(2p+2k+3)(2p+k+1)}-\frac{1}{(2p+2k+1)(k+1)}\right)\,.
\end{equation}
We further manipulate this double series as follows.  
First we use the partial fraction decomposition
\begin{align}\label{partialfraction}
&\frac{1}{(2p+2k+3)(2p+k+1)}-\frac{1}{(2p+2k+1)(k+1)}=\nonumber\\
&\qquad\qquad\frac{1}{2p-1}\left(
\frac{1}{p+k+3/2}-\frac{1}{2p+k+1}+\frac{1}{p+k+1/2}-\frac{1}{k+1}
\right)\,,
\end{align}
and then observe that the sum over $k$ can be expressed in terms of a sum
that gives a logarithm, and a finite sum. More precisely, we use 
\begin{equation}\label{digammaidentity}
\sum_{n=0}^\infty\left(\frac{1}{n+x}-\frac{1}{n+y}\right)=\psi(y)-\psi(x)\,
\end{equation}
where $\psi(x)$ is the Euler's digamma function that satisfies 
\begin{equation}\label{digammadefinition}
\psi(n)=-\gamma+\sum_{k=1}^{n-1}\frac{1}{k}\, ,\quad {\rm and}\quad
\psi(n+\tfrac{1}{2})=-\gamma-2\ln 2+\sum_{k=1}^{n}\frac{2}{2k-1}\,, \enspace n\in {\mathbb N} \, .
\end{equation}
 Using (\ref{digammaidentity}) and (\ref{partialfraction}) 
we obtain  
\begin{equation}\label{ccontrib1}
I_{-+}=2\pi^2\sum_{p=0}^\infty\frac{(-1)^p}{(2p+1)^2(2p-1)}
\left(4\ln2-\frac{2}{2p+1}+\sum_{k=1}^{2p}\frac{1}{k}-\sum_{k=1}^{p}\frac{4}{2k-1}\right)\,.
\end{equation}

\subsection{ Computation of $I_{++}^<$ and $I_{++}^>$}

In $I_{++}^<$ we start by using (\ref{J1}), (\ref{J2}).
The outcome has three terms, which we integrate by using
the expansion (\ref{exp}) and (\ref{GR_int}).
Combining the three terms one finds
\begin{equation}\label{I3++<intermediate}
I_{++}^<=4\pi^2\sum_{p=0}^{\infty}\frac{(-1)^p}{(2p+1)^2(2p+3)}\left(
\ln 2-\sum_{k=0}^p\frac{1}{2k+1}
\right)\, . 
\end{equation}
The integral $I_{++}^>$ is a bit special --- both integration variables 
show up in the argument of the logarithm. We compute the integration 
over $\phi_2$ (the angle of $u_2$) using (\ref{K}).
 Using (\ref{exp}) we perform
the integral over the angular variable of  $v_1$ that gives
\begin{equation}
I_{++}^>=\int_0^1d r \int_0^1 ds \, \frac{4\pi^2}{1-s^2}
\sum_{k=0}^\infty\frac{(-1)^k (rs)^{2k+2}}{(2k+1)^2}(s^{2k+1}-1)\,,
\end{equation}
where $r=|u_2|,\,s=|v_1|$. By means of \eqref{integral1}
we can perform the integrations over $s$ and $r$ and arrive at
\begin{equation}
I_{++}^>=4\pi^2\sum_{p=0}^\infty \frac{(-1)^p}{(2p+1)^2(2p+3)}\sum_{k=0}^\infty\left(
\frac{1}{4p+2k+4}-\frac{1}{2p+2k+3}
\right)\,.
\end{equation}
Using (\ref{digammaidentity}) to  sum over $k$ and the identity 
\begin{equation}
\tfrac{1}{2}(\,\psi(p+1+\tfrac{1}{2})-\psi(2p+2)\,)=-\ln{2}+\sum_{k=0}^p\frac{1}{2k+1}-\sum_{k=1}^{2p+1}\frac{1}{2k}\,.
\end{equation}
we find cancellations with terms in \eqref{I3++<intermediate}. We obtain a more compact expression for the sum
\begin{equation}\label{ccontrib2}
I_{++}^>+I_{++}^<=-2\pi^2\sum_{p=0}^\infty\frac{(-1)^p}{(2p+1)^2(2p+3)}\sum_{k=1}^{2p+1}\frac{1}{k}\,.
\end{equation}

\subsection{ Computation of $I_{--}^<$}

 Here we first perform the integration over the phase $\phi_2$ of $z_2$,
using \eqref{K}. The resulting integrand can be expanded for small $|u_1|$, using \eqref{exp},
and the angular variable $\phi_1$ of $u_1$ can be integrated out. This leaves us with
\begin{equation}
I_{--}^<=2\pi^2\int\limits_0^1\!\!ds\int\limits_0^1\!\!dr\sum_{p=0}^\infty\frac{(-1)^p(r^{4p+2}-r^{2p+1})
s^{2p+4-3\Delta}}{(2p+1)^2\,r(1-r^2)}\,.
\end{equation}
where $r=|u_1|$, $s=|z_2|$. The integral over $r$ is then
\begin{align}
&& \int\limits_0^1dr\,\frac{r^{4p+2}-r^{2p+1}}{r(1-r^2)}=\sum_{k=0}^\infty
\left(\frac{1}{4p+2k+2}-\frac{1}{2p+2k+1}\right)\nonumber\\
&& =\frac{1}{2}(\,\psi(p+\tfrac{1}{2})-\psi(2p+1)\,)
=\frac{1}{2}\left(-2\ln2+\sum_{k=0}^{p-1}\frac{2}{2k+1}-\sum_{k=1}^{2p}\frac{1}{k}\right)\,,
\end{align}
where we used \eqref{digammaidentity} and \eqref{digammadefinition}.
Taking the integral over $s$ and analytically continuing to $\Delta=2$ gives
\begin{equation}\label{ccontrib3}
I_{--}^<=2\pi^2\sum_{p=0}^\infty\frac{(-1)^p}{(2p+1)^2(2p-1)}
\left(-2\ln2+\sum_{k=0}^{p-1}\frac{2}{2k+1}-\sum_{k=1}^{2p}\frac{1}{k}\right)\,.
\end{equation}

\subsection{ Computation of $I_{+-}^{f1}+I_{--}^{>,f1}$}

These two integrals can be combined to give 
\be
I_{+-}^{f1}+I_{--}^{>,f1}= -\pi^2 I(3, \delta)
\ee
where $I(x,\delta) $ is defined in (\ref{Iint}).
Using  \eqref{Iintegralanswer}, (\ref{I3p}) we obtain
\be\label{ccontrib4}
I_{+-}^{f1}+I_{--}^>=3\ln2 I_3^p
+\lim_{\delta\rightarrow0}\frac{\pi^2 I(3,\delta)-I_3^p}{\delta}
=\pi^3\ln2+\frac{7}{2}\pi^3+2\pi^2 \,.
\ee

\subsection{ Computation of $I_{+-}^{f2}$}

We first use \eqref{tildeJ} then we use (\ref{exp})  to obtain
\begin{align}
I_{+-}^{f2}=\,&2\pi^2\sum_{p,k=0}^\infty\frac{(-1)^p}{(2p+1)^2}
\left(\frac{1}{(2p+2k+3)(2p+k+3)}-\frac{1}{(2p+2k+5)(k+1)}\right)\nonumber\\
&\quad+\,6\pi^2\sum_{p=0}^\infty\frac{(-1)^p}{(2p+1)^3(p+1)}\,.
\end{align}
In the first line we use partial fractions and the  digamma function identity \eqref{digammaidentity}
similarly as in the computation of $(I_3)_{-+}$ and find
\begin{align}\label{ccontrib5}
I_{+-}^{f2}=\,&12\pi^2\sum_{p=0}^\infty\frac{(-1)^p}{(2p+1)^2(2p-1)}
\left(4\ln 2+\sum_{k=1}^{2p+2}\frac{1}{k}-\frac{2}{2p+3}-\sum_{k=1}^{p+1}\frac{4}{2k-1}\right)\nonumber\\
&\quad+\,\pi^2\sum_{p=0}^\infty\frac{(-1)^p}{(2p+1)^3(p+1)}\,.
\end{align}

%

\subsection{ Computation of $I_{--}^{>,f2}$}

We have
\begin{equation}
I_{--}^{>,f2}=\pi\int_{\mathbb{D}^+}\frac{d^2z_1}{|z_1|^{3\Delta-2}}\,\ln\left|
\frac{1+iz_1}{1-iz_1}\right|\,\tilde{J}(e^{i\phi_1})\,.
\end{equation}
From the three summands in the factor $\tilde{J}(e^{i\phi_1})$ calculated in 
\eqref{tildeJ} we obtain three contributions to $(I_3)_{--}^{>,f2}$, 
which we integrate using the expansion \eqref{exp}. After  analytic 
continuation in $\Delta$ we obtain 
\begin{align}\label{ccontrib6}
I_{--}^{>,f2}=&\;4\pi^2\ln2\sum_{p=0}^\infty\frac{(-1)^p}{(2p+1)^2(2p-1)}
+4\sum_{p=0}^\infty\frac{(-1)^p}{(2 p+1)^3(2p-1)}\nonumber\\
&\;\quad-4\pi^2\sum_{p=0}^\infty\frac{(-1)^p}{(2 p+1)^2(2p-1)}\sum_{k=0}^p\frac{1}{2k+1}\,.
\end{align}

\subsection{The sum of all finite parts }

The sum of all integrals in Table~\ref{allfiniteintegrals} is now straightforward.
The results are contained in \eqref{ccontrib1}, \eqref{ccontrib2}, \eqref{ccontrib3}, \eqref{ccontrib4},
\eqref{ccontrib5}, and \eqref{ccontrib6} and yield
\begin{align}\label{I3fintermediate}
I_3^f=&\;8\pi^2\sum_{p=0}^\infty\frac{(-1)^p}{(2p-1)(2p+1)(2p+3)}
\left(2\ln2-\sum_{k=0}^p\frac{2}{2k+1}\right)\nonumber\\
&\quad+4\pi^2\sum_{p=0}^\infty\frac{(-1)^p\,(2p+7)}{(2p-1)(2p+1)^2(2p+3)^2}\\
&\qquad+\pi^3\ln2+\frac{7}{2}\pi^3+2\pi^2\,.\nonumber
\end{align}
For the single series we obtain the sums
\begin{equation}\label{cp1}
8\pi^2\sum_{p=0}^\infty\frac{(-1)^p}{(2p-1)(2p+1)(2p+3)}=-\pi^3\,,\\
\end{equation}
and
\begin{equation}\label{cp2}
4\pi^2\sum_{p=0}^\infty\frac{(-1)^p\,(2p+7)}{(2p-1)(2p+1)^2(2p+3)^2}=\frac{1}{2}\pi^3-3\pi^2-2\pi^2G \, 
\end{equation}
where $G$ is Catalan's constant.
 A little more work is required for the evaluation
of the double sum in (\ref{I3fintermediate}), which can be done by applying the same technique as
in the computation of $I_{(2)}$ in Appendix A, using integral (\ref{integral2}).
One obtains
\begin{equation}\label{cp3}
-8\pi^2\sum_{p=0}^\infty\frac{(-1)^p}{(2p-1)(2p+1)(2p+3)}\sum_{k=0}^p\frac{2}{2k+1}
=\frac{1}{2}\pi^3+\pi^2+2\pi^2 G +\pi^3\ln2\, .
\end{equation}
Combining \eqref{cp1},  \eqref{cp2}, and  \eqref{cp3}, \eqref{I3fintermediate} finally yields
\begin{equation}\label{I3fanswer}
I_3^f=\frac{9}{2}\pi^3 \,.
\end{equation}
Together with (\ref{I3p}) this gives the expansion
\be
i_3(\delta) = \frac{\pi^3}{\delta} + \frac{9}{2}\pi^3 + {\cal O}(\delta) \, .
\ee


\section{Useful integrals and series}
\setcounter{equation}{0} 
We frequently used the expansion 
\be\label{exp}
\ln\left| \frac{1+iz}{1-iz}\right| = i\sum_{k=0}^{\infty} \frac{(-1)^k}{2k+1}(z^{2k+1} - \bar z^{2k+1}) \, .
\ee
We use the following integral over the upper half disc:
\be \label{J1}
J(e^{i\phi}) = \int\limits_{{\mathbb D}^{+}} \frac{d^2 u} {|e^{i\phi} + u|^2} = -\frac{\pi}{2}\ln (\sin(\phi)) - \frac{\pi}{2}\ln 2 
+ [{\rm Cl}_2(\phi) -{\rm Cl}_2(\phi + \pi) ]
\ee
where 
\be 
{\rm Cl}_2(\phi) = \sum_{n=1}^{\infty} \frac{\sin(n\phi)}{n^2}
\ee
 stands for Clausen's integral (see e.g. \cite{Lewin}).
We have the following series representation:
\be\label{J2}
J(e^{i\phi}) = -\frac{\pi}{2}\ln(\sin(\phi)) - \frac{\pi}{2}\ln 2 + 2 \sum_{k=0}^{\infty} \frac{\sin((2k+1)\phi}{(2k+1)^2} \, . 
\ee
In addition to the integral $J$ we make use of the integral
\be
\tilde J(e^{i\phi}) = \int\limits_{{\mathbb D}^{+}} \frac{d^2 u}{|u|^2}\left( \frac{1} {|e^{i\phi} + u|^2} - 1\right) = 
-\frac{\pi}{2}\ln (\sin(\phi)) - \frac{\pi}{2}\ln 2 
-[ {\rm Cl}_2(\phi) -{\rm Cl}_2(\phi + \pi) ] \, ,
\ee
for which the series expansion reads 
\be\label{tildeJ}
\tilde J(e^{i\phi}) = -\frac{\pi}{2}\ln(\sin(\phi)) - \frac{\pi}{2}\ln 2 - 2 \sum_{k=0}^{\infty} \frac{\sin((2k+1)\phi}{(2k+1)^2} \, . 
\ee
Next we define 
\be \label{K}
K(u) = \int\limits_{0}^{\pi} \frac{d\varphi}{|e^{i\varphi} + u|^2} = \frac{1}{1-|u|^2}\Bigl[ 
\pi + i\ln\left(\frac{1+u}{1-u}\right) + i \ln\left(\frac{1-\bar u}{1+\bar u}\right) \Bigr]\, . 
\ee
Here are some expressions for 1-dimensional integrals we used:
\be\label{GR_int}
\int\limits_{0}^{\pi}\ln(\sin(\phi)) \sin[(2n+1)\phi] \, d\phi = \frac{2}{2n+1}\ln 2 + \frac{2}{(2n+1)^2} 
-\frac{4}{2n+1}\sum_{k=0}^{n}\frac{1}{2k+1}
\ee
(see \cite{GR}, formula GW (338)(3b)), and furthermore
\be \label{integral1}
\int\limits_{0}^{1}dx\, \ln(1-x^4){\rm arctanh}(x) x^{-3} = \frac{\pi}{4} - \frac{\pi^2}{16} - \frac{3}{2}\ln 2\, ,
\ee
\be \label{integral2} 
\int\limits_{0}^{1}\!\! dx\, \frac{\arctan(x)}{x}\left( \frac{x^2-1}{x^2+1}\right) = - \frac{\pi}{4}\ln 2\, .
\ee


\begin{thebibliography}{99}

\bibitem{Gaiotto} D. Gaiotto, 
{\it Domain walls for two-dimensional renormalization group flows},  
JHEP 12 (2012) 103; arXiv:1201.0767.

\bibitem{BPZ}
  A.~A.~Belavin, A.~M.~Polyakov and A.~B.~Zamolodchikov,
  {\it Infinite conformal symmetry in two-dimensional quantum field theory},
  Nucl.\ Phys.\ B {\bf 241} (1984) 333.
  
\bibitem{Schweigert_etal} J. Fuchs, I. Runkel, and C. Schweigert, 
{\it Twenty-five years of two-dimensional rational conformal field theory}, 
J. Math. Phys. {\bf 51} (2010) 015210; arXiv:0910.3145.

\bibitem{FQ} S. Fredenhagen and T. Quella, 
{\it Generalised permutation branes}, 
JHEP 0511:004, 2005; arXiv:hep-th/0509153.

\bibitem{BR} I. Brunner and D. Roggenkamp, 
{\it Defects and bulk perturbations of boundary Landau-Ginzburg orbifolds}, 
JHEP 04 (2008) 001; arXiv:0712.0188.

\bibitem{Oshikawa-Affleck}
  M.~Oshikawa and I.~Affleck,
  {\it Defect lines in the Ising model and boundary states on orbifolds},
  Phys.\ Rev.\ Lett.\  {\bf 77} (1996) 2604; arXiv:hep-th/9606177.

\bibitem{Gaberdiel_etal} S. Fredenhagen, M. R. Gaberdiel and C. A. Keller, 
{\it Bulk induced boundary perturbations}, 
J. Phys. {\bf A40}:F17, 2007;  arXiv:hep-th/0609034.

\bibitem{Zamolodchikov}
A. B. Zamolodchikov, 
{\it Renormalization group and perturbation theory about fixed points in two-dimensional field theory}, 
Sov. J. Nucl. Phys. {\bf 46} (1987) 1090.

\bibitem{GKSC} M. R. Gaberdiel, A. Konechny and C. Schmidt-Colinet, 
{\it Conformal perturbation theory beyond the leading order}, 
J.Phys. {\bf A42} (2009) 105402; arXiv:0811.3149.

\bibitem{AL1} I. Affleck and A. W. W. Ludwig, 
{\it Universal non integer `ground state degeneracy' in critical quantum systems}, 
Phys. Rev. Lett. {\bf 67} (1991) 161.

\bibitem{CL}
A. W. W. Ludwig and J. Cardy, 
{\it Perturbative evaluation of the conformal anomaly at new critical points with applications to random systems}, 
Nucl. Phys. {\bf B285} (1987) 687.

\bibitem{AL2} I. Affleck and A. W. W. Ludwig, 
{\it Exact conformal field theory results on the multi-channel
Kondo effect: Single-fermion Green's function, self-energy and resistivity}, 
Phys. Rev. {\bf B48} (1993) 7297.

\bibitem{FK} D. Friedan and A. Konechny, 
{\it Boundary entropy of one-dimensional quantum systems at low temperature}, 
Phys. Rev. Lett. {\bf 93} (2004) 030402; arXiv:hep-th/0312197.

\bibitem{Florens_Rosch} S. Florens and A. Rosch, 
{\it Climbing the entropy barrier: driving the single -- towards the multi critical Kondo effect by a 
weak Coulomb blockade of the leads}, Phys. Rev. Lett. {\bf 92} (2004) 216601.

\bibitem{DF} 
  V.S. Dotsenko and V.A. Fateev, 
  {\it Operator algebra of two-dimensional conformal theories with central charge $c\le 1$}, Phys. Lett {\bf B154} (1985) 291; 
  V.~S.~Dotsenko and V.~A.~Fateev, {\it Four point correlation functions and the operator algebra in the two-dimensional conformal 
  invariant theories with the central charge $c\le 1$}, 
  Nucl.\ Phys.\ B {\bf 251} (1985) 691.

\bibitem{Lassig} 
M. L\"assig, 
{\it Geometry of the renormalization group with an application in two dimensions}, 
Nucl. Phys. {\bf B334} (1990) 652. 

\bibitem{Constantinescu-Flume}
F. Constantinescu and  R. Flume, 
{\it Perturbation theory around two-dimensional critical systems through holomorphic decomposition}, 
J. Phys. {\bf A23} (1990) 2971.

\bibitem{CSS} C. Crnkovi\'c, G.M. Sotkov and M. Stanishkov, 
{\it Renormalization group flow for general $SU(2)$ coset models}, 
Phys. Lett. {\bf B226} (1989) 297.

\bibitem{Poghossian^2} A. Poghosyan and H. Poghosyan, 
{\it Mixing with descendant fields in perturbed minimal CFT models}, 
JHEP {\bf 10} (2013)131; arXiv:1305.6066.  

\bibitem{Poghossian} R. Poghossian, 
{\it Two Dimensional Renormalization Group Flows in Next to Leading Order}, 
JHEP 01 (2014) 167; arXiv:1211.3665. 

\bibitem{Lewin} L. Lewin, 
{\it Polylogarithms and associated functions}, 
Elsevier Science Ltd (1981).

\bibitem{GR} L. S. Gradshteyn and I. M. Ryzhik, 
{\it Table of integrals, series, and products}, Academic Press (2000).

\end{thebibliography}
\end{document}